\documentstyle[epsfig,epsf]{article}
\def\Journal#1#2#3#4{{#1}{\bf #2} (#4) #3}

\def\NCC{{\em Nuovo Cimento} {\bf C}}
\def\NAT{\em Nature~}
\def\LNC{\em Lett. Nuovo Cimento~}
\def\NIMA{{\em Nucl. Instr. Meth.} {\bf A}}
\def\ASS{\em Astrophys. Space Sci.~}
\def\PSS{\em Planet Space Sci.~}
\def\PLB{{\em Phys. Lett.} {\bf B}}

\def\PRD{{\em Phys. Rev.} {\bf D}}

\def\EPJC{{\em Eur. Phys. J.} {\bf C}}
\def\AP{\em Astrop. Phys.~}

\def\EL{\em Europhys. Lett.~}
\def\APJ{\em Ap. J.~}

\begin{document}

\begin{center}
{\Large {\bf Time variations in the deep underground muon flux measured 
by MACRO}}
\end{center}

\vskip .7 cm

\begin{center}
Y. Becherini$^1$, S. Cecchini$^{1,5}$, M. Cozzi$^1$, H. 
Dekhissi$^{1,2}$, J. Derkaoui$^2$, G. Giacomelli$^1$,  
 M. Giorgini$^1$, F. Maaroufi$^2$, G. Mandrioli$^1$, A. Margiotta$^1$, S. 
Manzoor$^{1,3}$, A. Moussa$^2$, L. Patrizii$^1$, V. Popa$^{1,4}$, M. 
Sioli$^1$, G. Sirri$^1$, M. Spurio$^1$ and V. Togo$^1$
 \par~\par

{\it 
(1) Dept of Physics, Univ. of Bologna and INFN Bologna, 40127 Bologna, 
 Italy \\ 
(2) LPTP, Faculty of Sciences, University Mohamed 1$^{st}$, B.P. 424, Oujda, 
 Morocco\\
(3) PRD, PINSTECH, P.O. Nilore, Islamabad, Pakistan  \\
(4) ISS, 77125 Bucharest-Magurele, Romania\\
(5) INAF/IASF Sez. di Bologna, 40129 Bologna, Italy \\}

\par~\par

Talk given by S. Cecchini at the $29^{th}$ ICRC, Pune, India, 3-10 August 2005

\vskip .7 cm
{\large \bf Abstract}\par
\end{center}

{\normalsize 
More than 30 million of high-energy muons collected with the MACRO detector 
at the underground Gran Sasso Laboratory have been used to search for flux 
variations of different natures. Two kinds of studies were carried 
out: search for periodic variations and for the occurrence of clusters 
of events. Different analysis methods, including Lomb-Scargle spectral analysis and Scan Test statistics have been applied to the data.}

\vspace{5mm}

\section{Introduction}\label{sec:intro}
The high energy muon events collected by the MACRO apparatus at the 
average depth of 3600 m.w.e. represent one of the most extensive records 
of such kind of data. The series of these high-energy muons can be to 
search for time variations of periodic and of stochastic characters, as 
it was done extensively by using arrival times of EAS \cite{first}. These 
variations in the underground muon flux may be due to different causes 
of galactic, solar and terrestrial origin. The common problem for this 
type of searches is to determine whether an observed effect has occurred 
by chance of if it signals a departure from a pure random muon arrival. \par
MACRO was a multipurpose modular apparatus with 6 supermodules 
with scintillator detectors, limited streamer tubes and nuclear track 
detectors \cite{MACRO}, and studied atmospheric neutrinos \cite{neu}, aspects 
of CR physics and astrophysics \cite{crmu}, searched for GUT Magnetic 
Monopoles and other exotica \cite{MM}. 
Some interruptions of different kinds occurred during data taking, either 
randomly (e.g. power outages), or regularly (e.g. maintenance), so 
appropriate statistical methods have to be applied and particular 
care should be used in choosing periods of stationary conditions. \par
In the following we discuss the results of the searches for periodic 
variations and for time clustering of muon events.

\section{Periodicity search. Spectrum analyses}\label{sec:periodicity}

For this analysis we considered data recorded by the streamer tube 
system in the time interval November 1991-May 2000 and selected the 
data with the following criteria: \par
- run duration longer than 1 hour; \par
- streamer tube efficiencies of wires and strips larger than 90\% 
and 70\%, respectively, for each module; \par
- all 6 super-modules in acquisition; \par
- acquisition dead time smaller than 2.5\% for the whole detector. \par

The total number of runs surviving these cuts was 6920 for a total number 
of $3.5 \cdot 10^7$ muon events. 

The Fourier amplitude spectrum analysis is a powerful technique that allows 
a blind search for regular/persistent fluctuations in time 
series \cite{attolini}. Such a technique, however, requires the input data 
to be sampled at evenly spaced intervals; data gaps of variable length 
and occurring randomly in the serie produce spurious contributions to 
the power that can mimic the presence of a periodicity. The Lomb-Scargle 
method \cite{lomb} has been developed to mitigate this effect even in the 
case of very long data series. Moreover, as indicated in \cite{prob}, it 
allows to evaluate the significance of the ``peaks'' (signal) 
with respect to a null hypothesis.
The muon events were binned in 15 min time intervals and bins deviating 
by more than $3 \sigma$ from the monthly average rate were discarded. The 
total number of time bins used was 160242 corresponding to 58\% of the 
whole sample.

\begin{figure}
\begin{center}
\mbox{\epsfig{figure=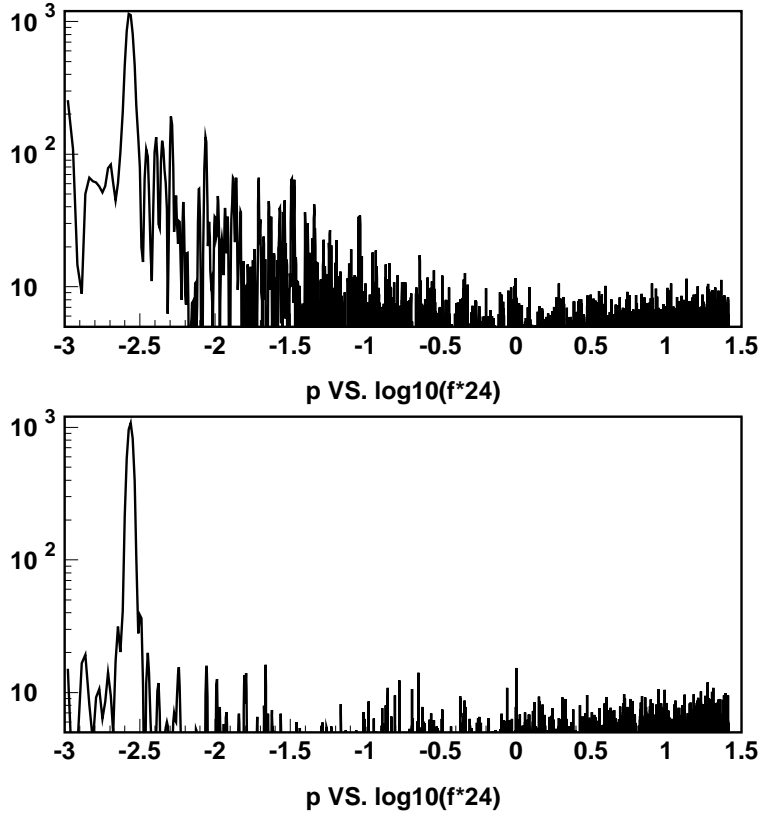,height=12cm}}
\caption {Lomb power as a function of the Log$_{10}$ of the frequency
[days$^{-1}$] for experimental data (upper panel). Note the high peak 
at $\sim -2.56$ corresponding 365 days (seasonal flux variation). In 
the lower panel the results
of a Monte Carlo simulation having the same noise level of the real data 
with seasonal, solar diurnal and sidereal waves \cite{MACanis} added.}
\label{fig:1}
\end{center}
\end{figure}

\begin{figure}
 \begin{center}
\mbox{\epsfig{figure=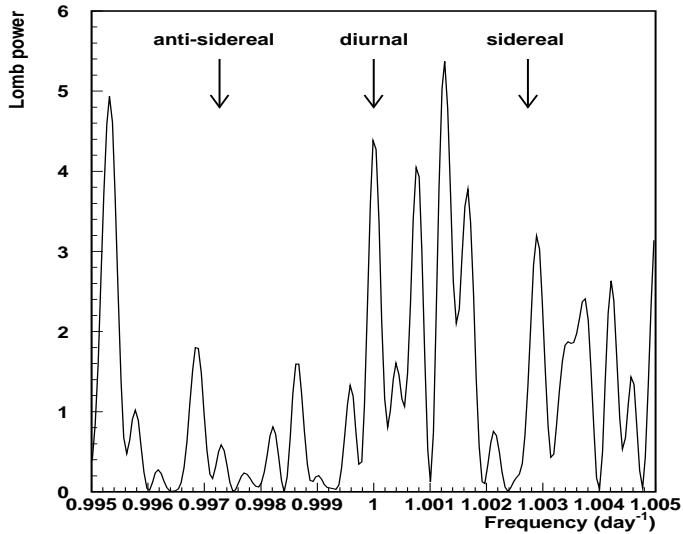,height=8cm,width=10cm}}
\caption{The frequency region around the solar diurnal wave. The arrows 
mark its position and the sidereal and anti-sidereal peaks.}
\label{fig:2}
\end{center}
\end{figure}

The results of our analysis are shown in Fig. \ref{fig:1}. We compare 
the spectrum 
obtained for the real data with a Monte Carlo simulation having the same noise level and time intervals distributed 
according to the sequence of the original series.
The seasonal, solar diurnal and sidereal waves \cite{MACanis} were also 
added in the serie. The 
spectrum of real data shows a power distribution similar to what 
observed in other cosmic ray data series \cite{attolini}, i.e. a low 
frequency spectrum whose power decreases with frequency$^{-2}$. The most 
striking feature of the spectrum is the large peak at $\sim -2.56$ 
corresponding to the seasonal flux variation.
Fig. \ref{fig:2} shows a frequency region around the solar diurnal frequency 
where we have also indicated the frequencies corresponding to the sidereal 
and anti-sidereal waves. The peak at frequency 1 day$^{-1}$ has a statistical 
significance of $\sim 2.3 \sigma$; the statistical significance assuming
an oscillatory behaviour is  $\sim 3.4 \sigma$. 
Note that peaks corresponding to the diurnal and sidereal variations
are observed, but peaks of similar size 
(or even larger) are present elsewhere in the spectrum. The claim that 
the sidereal and solar diurnal waves are real is based upon its occurrence 
at a frequency of ``a priori'' interest and on the stability of its 
amplitude and phase with time. We find that the amplitudes and the 
probabilities for the null hypothesis computed by the two methods are 
in fair agreement with the ones obtained using a standard ``folding'' 
method \cite{MACanis}.

\section{Burst search: time interval distribution}\label{sec:burst1}
\begin{figure}
 \begin{center}
\mbox{\epsfig{figure=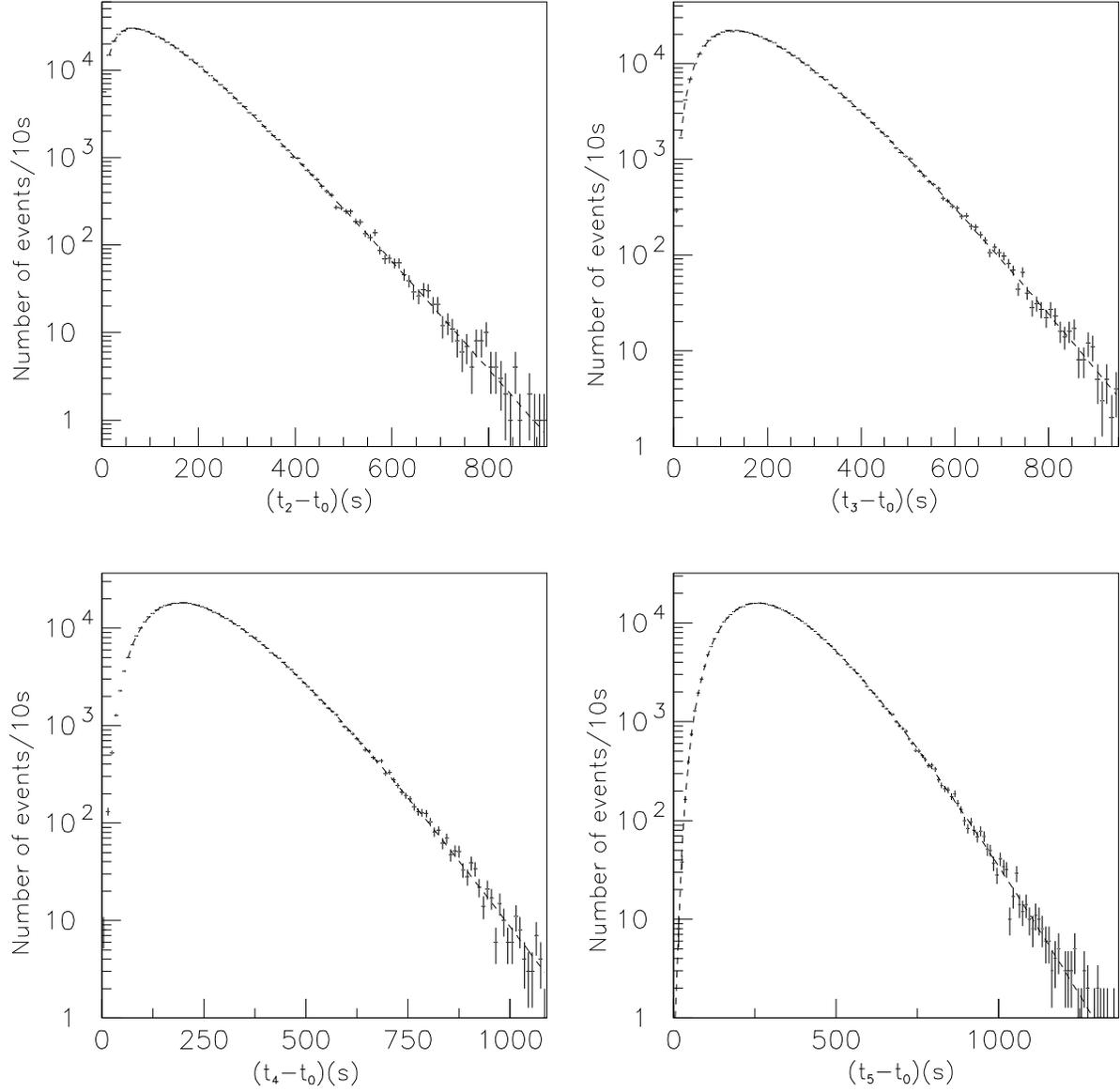,height=18cm}}
\caption{ Time interval distributions (from top left to bottom left 
clockwise) for t$_2$-t$_0$, t$_3$-t$_0$, t$_5$-t$_0$ and t$_4$-t$_0$, 
 respectively, for a cone of arrival directions with declination 
$25^\circ < \delta < 50^\circ$. The dashed lines represent the fits 
to the Erlangen gamma function of order 2, 3, 5, 4, respectively 
(see \cite{erlangen}).}
\label{fig:3}
\end{center}
\end{figure}

The first method used in searching for correlations in the arrival times 
was the study of the time interval distribution. For each muon arriving 
at time t${_0}$ we calculated the distribution of the time interval 
elapsed between the first muon t${_0}$ and 
the next five muons: t${_i}$-t${_0}$, i=1,...,5, see Fig. \ref{fig:3}. 
 A complete analysis was published in 
\cite{erlangen}. Here we report the results for the direction bands 
with $0^\circ <$ RA$< 360^\circ$ and 25$<$ decl $<50^\circ$ that 
include the Cyg X-3 
region. The experimental distributions show some deviations 
from the muon random arrival expectation. The probability computed using 
the Kolmogorov-Smirnov test shows some disagreement (prob=0.38) but the 
available statistics is too poor to reach clear conclusions.

\section{Burst search: scan statistics}\label{sec:burst2}
\begin{figure}
 \begin{center}
\mbox{\epsfig{figure=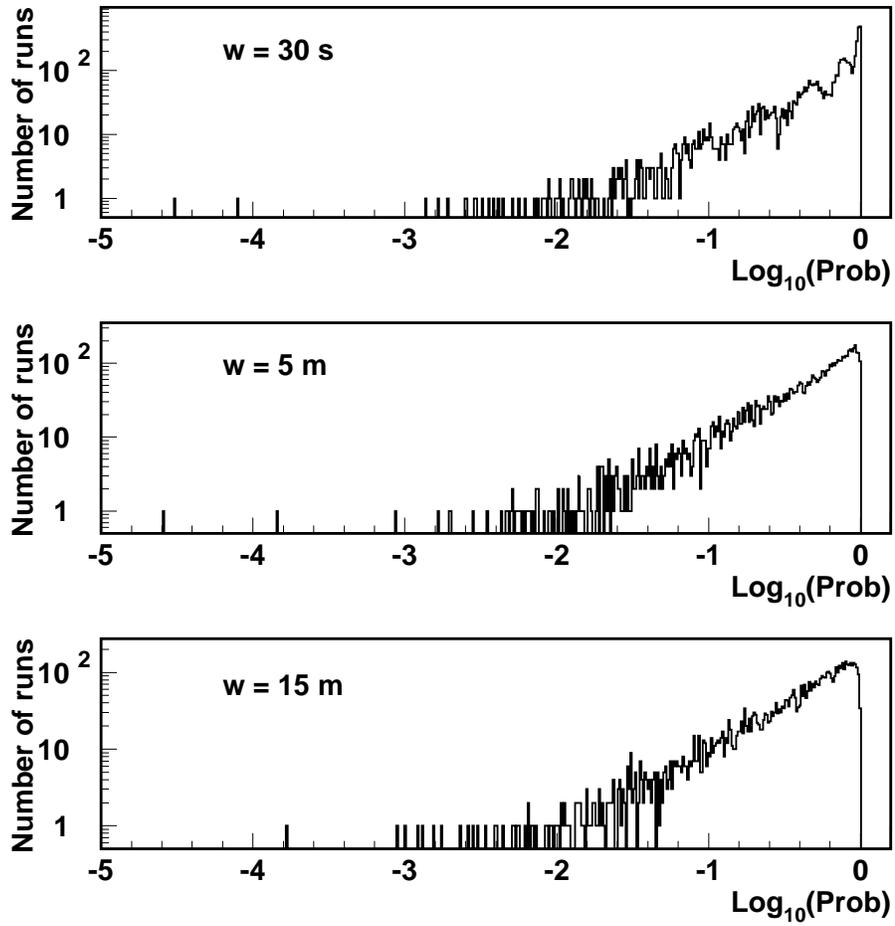,height=14cm}}
\caption{Scan statistics probability distributions for the selected runs. 
In the upper panel a time window $w=30$ s was used; $w=5$ min in the central 
panel and  $w=15$ min in the bottom panel.}
\label{fig:4}
\end{center}
\end{figure}

Scan statistics is a powerful method to search for bursts of events.
It is a bin-free method and it provides unbiased results when data are
analysed (see \cite{scan} and references therein). 
We used scan statistics in the following way: for each 
run $i$, let $[A_i,B_i]$ be the time interval ranging from the start and 
the end of the run. We open a ``time window'' of fixed length $w$ and scan 
the interval $[A_i,B_i]$ counting the number of events falling inside 
$w$. $k_i$ is the maximum number of events 
recorded during the scan. Finally, for each run, we compute 
the probability $P_i$ that 
a statistical fluctuation would produce a burst of events as large as $k_i$.
The only  choice is the size of $w$. We tried different 
sizes ($w=30$ s, 5 min and 15 min) and for each of them we analysed the 
probability distribution $P_i$, i=1, N$_{run}$. In Fig. \ref{fig:4} 
we show the probability distribution for the 6113 runs surviving our cuts:
$w=30$ s above, 5 min at the centre and 15 min below.
No significant deviations from the null 
hypothesis is found; we also inspected unusual runs with  
probabilities smaller than $5\cdot 10^{-4}$ and we found that the ``bumps'' 
of events were concentrated near the beginning or end of the runs.

\section{Conclusions}\label{sec:conclu}
We analysed the time series of MACRO muons using two complementary 
approaches: search for periodicities and search for bursts of events. The
Lomb-Scargle method was used in the first case, scan statistics in the 
second. The two techniques complete early analyses performed with ``folding''
methods in searching for periodicities and time differences for burst 
events. No deviations from the expected distributions were found.

\section{Acknowledgements}
We thank the members of the MACRO Collaboration and the personnel 
of the LNGS for their cooperation. H.D. and S.M. thank the Abdus Salam 
ICTP TRIL Programme for providing fellowships. H.D., J.D., G.G., F.M. 
and A. Moussa thank the collaboration between the Universities of 
Bologna and Mohamed $1^{st}$ of Oujda.

\end{document}